\documentclass[runningheads,a4paper]{llncs}

\usepackage{amssymb}
\usepackage{graphicx}
\usepackage{algorithm}
\usepackage{algpseudocode}
\usepackage{pifont}

\usepackage{comment}
\usepackage{graphicx}
\usepackage{epstopdf} 
\usepackage{amsmath}           
\usepackage{amssymb}
\usepackage{thmtools,thm-restate}         
\usepackage{multirow}
\usepackage{multicol}
\usepackage{array,booktabs}
\usepackage{threeparttable}
\usepackage{subfig}
\usepackage{color}

\usepackage{url}
\usepackage{hyperref}  
\hypersetup{
	colorlinks = true,         
	citecolor = blue,
	linkcolor = red,          
	urlcolor = blue,           
}

\newcolumntype{P}[1]{>{\centering\arraybackslash}p{#1}}
\newcolumntype{M}[1]{>{\centering\arraybackslash}m{#1}}

\begin{document}

\mainmatter  

\title{Comments on ``Defeating HaTCh: Building Malicious IP Cores''}

\titlerunning{Comments on "Defeating HaTCh: Building Malicious IP Cores"}

%
%

\author{Syed Kamran Haider \and Chenglu Jin \and Marten van Dijk}

%


\institute{Department of Electrical and Computer Engineering\\ University of Connecticut, Storrs, CT, 06269, USA}

%
%

\maketitle

\section{Introduction}

Recently, Haider \emph{et al.} introduced the first rigorous hardware Trojan detection algorithm called HaTCh~\cite{haider2017advancing}. The foundation of HaTCh is a formal framework of hardware Trojan design introduced in~\cite{haider2017advancing_MWSCAS}, which formally characterizes all the hardware Trojans based on its properties.  

However, Bhardwaj \emph{et al.} recently published one paper ``Defeating HaTCh: Building Malicious IP Cores''~\cite{bhardwaj2017defeating}, which \textit{incorrectly} claims that their newly designed hardware Trojan can evade the detection by HaTCh. The coding scheme used for transmitting secret keys in~\cite{bhardwaj2017defeating}  is interesting and valuable  to the community, but \textbf{the claim that this new Trojan can defeat HaTCh is incorrect}, due to the authors' misunderstanding of HaTCh~\cite{haider2017advancing} and its fundamental definitional framework~\cite{haider2017advancing_MWSCAS}. 

\section{Formal Framework for Hardware Trojans}

In~\cite{haider2017advancing_MWSCAS}, the class of hardware Trojans that the authors formally characterized is called $H_D$, which is referred to the class of \textbf{trigger-based pre-silicon} hardware Trojans \textbf{embedded in a deterministic function (IP core)}, and its malicious payload is \textbf{only delivered by Standard I/O channels}. In $H_D$ class, the authors further discovered four advanced properties ($d$, $t$, $\alpha$ and $l$) to formally characterize all possible hardware Trojans in $H_D$. The readers are referred  to~\cite{haider2017advancing_MWSCAS} and its extended version~\cite{haider2016advancing} for a detailed discussion of these four properties.

Based on this formal definitional framework, Haider \textit{et al.} proposed one hardware Trojan detection algorithm, called HaTCh, that offers \textit{negligible false negative rate} and \textit{controllable false positive rate} for the detection of \textbf{all possible hardware Trojans in} $H_D$~\cite{haider2017advancing}. The readers are referred to~\cite{haider2017advancing} for a detailed algorithm and formal proofs. 

\section{Detailed Hardware Trojan Design in~\cite{bhardwaj2017defeating}}

The hardware Trojan design in~\cite{bhardwaj2017defeating} is based on an AES (Advanced Encryption Standard~\cite{pub2001197}) core and a PRSG (Pseudorandom bit Sequence Generator). The authors argued that in some cases, in order to provide traffic flow confidentiality as suggested in~\cite{kiraly2008traffic}, the output port of an AES core needs to keep outputting random bit strings even when the AES core is not working. The authors introduced a coding method to embed the secret key into the random string sent out during the idle time to establish a covert channel with an eavesdropping adversary. Since the adversary knows how the key is encoded, he/she is able to decode the random string and retrieve the key. The authors also ran some statistical analysis to show that the random string generated by their hardware Trojan is indistinguishable from a ``real'' random string, so they choose to keep their hardware Trojan \textbf{always on}, in other words, this Trojan does not have a trigger. This is why they claim they can evade the detection of HaTCh and defeat HaTCh, but actually \textbf{this hardware Trojan design is not in $H_D$ class, because it is not a trigger-based hardware Trojan.} Thus, their final claim of defeating HaTCh is incorrect, because HaTCh is designed for detecting all possible Trojans in $H_D$ class. \footnote{An always-on hardware Trojan which leaks information via a standard digital channel can be deterministic, but as explained in~\cite{haider2017advancing} it will not even pass functional testing.  Section~\ref{sec:update} explains that in fact the hardware Trojan is not always on but it is still not in $H_D$ because it is embedded in a non-deterministic IP core.}      

\section{Common Misunderstandings about HaTCh}

It is usually very easy for the readers of~\cite{haider2017advancing,haider2016advancing} to miss some crucial requirements for using HaTCh. Let us list some important properties of hardware Trojans to fall within the scope of HaTCh as follows:

\begin{enumerate}
\item The hardware Trojans need to be trigger-based, so the Trojans in~\cite{bhardwaj2017defeating,lin2009trojan} are not in $H_D$ class, and therefore HaTCh cannot provide security guarantees. 

\item The malicious payload should only be delivered by standard I/O channel, so the side channel Trojans, e.g.~\cite{lin2009trojan,ender2017first}, are out of the scope of HaTCh.

\item HaTCh is designed for pre-silicon Trojans, so all the Trojans inserted during fabrication, like~\cite{becker2013stealthy,yang2016a2}, will not be detected by HaTCh. 

\item HaTCh requires the original IP core to be a deterministic function for its functional testing, and therefore any IP core has non-deterministic function (True Random Number Generator) needs to test its deterministic part separately. 

\item Since HaTCh algorithm takes the four advanced properties ($d$, $t$, $\alpha$ and $l$) of hardware Trojans as inputs, so a HaTCh instance with a given set of these four parameters can provably detect all possible Trojans in the class of $H_{d,t,\alpha,l}$.   
\end{enumerate}

Another common misunderstanding about HaTCh is that the way how a Trojan is triggered will affect the detection capability of  HaTCh. This is not correct. As long as there is a \textbf{trigger activation condition} that only appears when the Trojan is triggered, HaTCh is able to detect it. Thus, a clock glitch triggered Trojan~\cite{ali2011multi} or a sensor triggered Trojan~\cite{ng2015integrated} both fall into the category of $H_D$, and therefore will lead to detection by HaTCh. 

\section{Conclusion}

HaTCh~\cite{haider2017advancing} and its definitional framework~\cite{haider2016advancing} are the first formal treatment of hardware Trojan design and detection. They are supported by formal analysis and rigorous proofs, so the readers are suggested to read and understand the theory and proof in their study.  

\section{Recent Updates on~\today}\label{sec:update}

Recently, Bhardwaj \emph{et al.} uploaded one new article ``Validating the Claim - Defeating HaTCh : Building Malicious IP cores''~\cite{bhardwaj2018validating} on 9/18/2018. According to the arguments presented in~\cite{bhardwaj2018validating}, it shows that the authors still failed to fully understand the formal definitional framework in~\cite{haider2016advancing} that our detection algorithm HaTCh~\cite{haider2017advancing} is based on. 

\subsection{The Claim in~\cite{bhardwaj2018validating}}
The authors claim that their hardware Trojans introduced in~\cite{bhardwaj2017defeating} is in the class of $H_D$, and it is not an always-on hardware Trojan. Its trigger mechanism is embedded in the AES core, because when the AES core is not encrypting anything, the hardware Trojan is active and leaking secret key. 

\subsection{Our Claim}
\textbf{The hardware Trojan introduced in~\cite{bhardwaj2017defeating} is not in $H_D$.} We agree with the claim that one can consider the trigger mechanism is in the AES core, if one considers the AES core, PRSG and the MUX jointly as a single IP core. However, there is a simple fact that shows why the hardware Trojan in~\cite{bhardwaj2018validating} is not in $H_D$. \textbf{The whole IP core, comprising of AES, PRSG and MUX, does not have a deterministic behaviour.} 

Let us first revisit the definition of $H_D$ in~\cite{haider2016advancing}. 

\begin{definition}
\cite{haider2016advancing} \boldmath{$H_D$} Trojans are the ones which are: 
\begin{enumerate}
\item Embedded in an IP core whose output is a function of only its input -- i.e. the logical functionality of the IP core is deterministic, and
\item The algorithmic specification of the IP core can exactly predict the IP core behavior.
\end{enumerate}
\end{definition}

Let us see the reason why a detection algorithm will not be able to catch the abnormal behaviour of the hardware Trojan in ~\cite{bhardwaj2017defeating}. 

Quoting from~\cite{bhardwaj2017defeating}: ``The key bits are randomized before transmission by the Trojan, so to a benign observer the flow of output bit patterns appears random, much like the cipher text and/or the PRSG bits, however an eavesdropper with prior knowledge of the proposed Trojan design can extract the encryption key bits.''

Hardware Trojan class $H_D$ requires a specification to exactly predict the IP core behavior, so if a benign observer (or a detection algorithm like HaTCh) sees an output does not match the specification, it should have already detected this Trojan, let alone when a random-looking output is observed. Thus, the reason why the Trojan proposed in~\cite{bhardwaj2017defeating} can evade detection relies on the fact that the specification of the large IP core allows random outputs. This means that the IP core, compromising of AES, PSRG and MUX, is not a deterministic function, so any hardware Trojan embedded in this IP core is not in $H_D$.

Moreover, the analysis of the four properties ($d$, $t$, $\alpha$ and $l$) of their hardware Trojan in~\cite{bhardwaj2018validating} is incorrect as well. The authors only show a few vague explanations to argue that the parameters of their Trojan are very large, but given the precise mathematical definitions of these four properties in~\cite{haider2016advancing}, one can precisely calculate the values of these four properties, instead of saying ``they are extremely large''. In particular, the trigger signal dimension of the Trojan in~\cite{bhardwaj2017defeating} is very likely to be 1, instead of being very large. In fact, one can determine whether the Trojan is active or not by checking the selection wire (single wire) of the 2-to-1 MUX, that chooses to output either the AES output (when AES is running and the Trojan is deactivated) or PRSG output (when AES is not running and therefore the Trojan is active). 

\subsection{Summary} We strongly encourage the authors of~\cite{bhardwaj2017defeating,bhardwaj2018validating} to study the full version of our theory paper~\cite{haider2016advancing}, and understand the formal rigorous definitions in~\cite{haider2016advancing}, especially Definition 6 to 11. 

We also would like to emphasize that HaTCh framework is formally and rigorously proved in~\cite{haider2017advancing}. So, a claim like ``defeating HaTCh'' is equivalent to finding a mistake in our proofs or designing a hardware Trojan does not satisfy the statistical assumptions that our claims and proofs are based on. However, notice that a claim like ``defeating HaTCh'' does not mean designing a hardware Trojan not in our characterized hardware Trojan classes. 

%

\bibliographystyle{splncs03}

\bibliography{ref}

\end{document}